%% file: main.tex
\documentclass[conference]{IEEEtran}
\IEEEoverridecommandlockouts

\usepackage{cite}
\usepackage{amsmath,amssymb,amsfonts}
\usepackage{algorithmic}
\usepackage{graphicx}
\usepackage{textcomp}
\usepackage{xcolor}

\def\BibTeX{{\rm B\kern-.05em{\sc i\kern-.025em b}\kern-.08em
    T\kern-.1667em\lower.7ex\hbox{E}\kern-.125emX}}

\usepackage{subfigure}
\usepackage{tabularray}
\usepackage{multirow}
\usepackage[hidelinks]{hyperref}
\usepackage{enumitem}
\usepackage{stfloats}
\usepackage{tablefootnote}
\usepackage{placeins}
\usepackage{color,soul}
\usepackage{threeparttable}
\usepackage[percent]{overpic}
\usepackage{circuitikz}

\title{A Dry-Contact Ear-EEG System With Continuous Electrode-Skin Impedance Mismatch Monitoring for Motion Artifact Cancellation Using DRL Stimulus}

\author{
\IEEEauthorblockN{
Lohan Atapattu\textsuperscript{\textdagger1},
Sajitha Madugalle\textsuperscript{\textdagger1},
Imasha Nethmal\textsuperscript{\textdagger1},
Erandee Jayathilaka\textsuperscript{\textdagger1},\\
Avishka Herath\textsuperscript{1},
Kithmin Wickremasinghe\textsuperscript{2},
Simon L. Kappel\textsuperscript{3},
Nilan Udayanga\textsuperscript{4},
Chamira U. S. Edussooriya\textsuperscript{1}
}

\vspace{0.3cm}

\IEEEauthorblockA{
\textsuperscript{1}Department of Electronic and Telecommunication Engineering, University of Moratuwa, Sri Lanka.\\
\textsuperscript{2}Department of Electrical and Computer Engineering, University of British Columbia, Vancouver, Canada.\\
\textsuperscript{3}Department of Electrical and Computer Engineering, Aarhus University, Denmark.\\
\textsuperscript{4}Cirtec Medical Corporation, USA.\\
\vspace{-0.7cm}
\textsuperscript{\textdagger}These authors contributed equally to the work.
}
\thanks{This is the author's version of the article that has been accepted for
presentation at the 2026 IEEE Biomedical Circuits and Systems Conference
(BioCAS 2026).

\copyright~2026 IEEE. Personal use of this material is permitted.
Permission from IEEE must be obtained for all other uses, in any current or
future media, including reprinting/republishing this material for advertising
or promotional purposes, creating new collective works, for resale or
redistribution to servers or lists, or reuse of any copyrighted component of this work in other works.}
}

\begin{document}

\maketitle

\begin{abstract}
Dry-contact ear-electroencephalography (Ear-EEG) enables wearable neural monitoring. However, motion induced electrode-skin impedance (ESI) mismatches between electrodes can severely degrade signal quality. To the best of our knowledge, this paper presents the first proof-of-concept dry-contact Ear-EEG system that uses a driven-right-leg (DRL) stimulus for continuous ESI mismatch monitoring, enabling online adaptive motion artifact cancellation. 
A 1 kHz sinusoidal stimulus is injected through the DRL electrode. The resulting response to the injected carrier is separated from the EEG using bandpass filtering and demodulation, and then used to extract the ESI mismatch information as the reference input for a normalized least-mean-square adaptive filter followed by a Hampel filtering stage. To evaluate artifact suppression and preservation of neural activity, alpha-band EEG activity was analyzed involving four healthy participants performing head nodding, electrode tapping, and jaw clenching. The system achieved artifact power reductions of 6.5, 12.6, and 9.0 dB (77.6\%, 92.8\%, and 86.4\%, respectively) while alpha-band modulation remained clearly observable after processing. This demonstrates the feasibility of DRL-stimulus-based ESI mismatch monitoring for motion artifact cancellation in wearable dry-contact Ear-EEG.

\end{abstract}

\vspace{-0.4cm}
\section{Introduction}

Dry-contact ear-electroencephalography (Ear-EEG) enables unobtrusive neural monitoring in daily life through compact ear-worn or hearing-aid-like devices suitable for long-term recording \cite{mikkelsen_eeg_2015}. This form factor supports applications in sleep monitoring, cognitive-state assessment, neurofeedback, and brain computer interfaces \cite{mikkelsen2019earEEGsleep,ZIBRANDTSEN20172454}. However, reliable acquisition during natural movement remains challenging because motion induced changes at the electrode-skin interface can severely degrade EEG signal quality. These artifacts arise primarily from head movement, jaw motion, and contact-pressure variations \cite{kappel2017physiological}, and their spectra overlap with key EEG bands, including delta, theta, alpha, and beta. Consequently, frequency domain filtering may remove neural information together with artifacts \cite{giangrande_motion_2024}. Although signal decomposition, wavelet-based methods, and deep learning have shown promising artifact removal performance \cite{10708999,s22082948,stalin2021machine,inproceedings}, many require high computational cost, multiple channels \cite{azemi2023biosignal}, large datasets, or offline processing, limiting their applicability to low power wearable Ear-EEG systems.

To address these limitations, adaptive filtering has been used for artifact reduction in wearable EEG systems \cite{10.3389/fnins.2021.611962,10286273}. In dry-contact Ear-EEG, electrode-skin impedance (ESI) variations are well suited as a reference for motion artifact reduction because they directly reflect motion induced changes at the electrode-skin interface.  However, most existing ESI monitoring methods inject the measurement current through the sensing electrodes, which requires high-input-impedance, low-noise driver circuits under high contact impedance conditions and can potentially degrade the common-mode rejection performance of the analog front-end (AFE) \cite{muller}. In contrast, Degen \textit{et al.} demonstrated continuous ESI mismatch monitoring for electrocardiogram (ECG) acquisition by injecting a low amplitude, high frequency excitation signal through the driven-right-leg (DRL) path and extracting the impedance related response during biosignal acquisition \cite{degen2008continuous}.

To the best of our knowledge, this work presents the first proof-of-concept dry-contact Ear-EEG platform that extends the DRL based ESI monitoring approach of Degen \textit{et al.}  to EEG acquisition and enables online motion artifact cancellation (MAC) on an embedded microcontroller. A low amplitude 1 kHz sinusoidal stimulus is injected through the DRL electrode, and the carrier response is demodulated to derive an ESI mismatch correlated artifact reference. This reference drives a normalized least-mean-square (NLMS) adaptive filter with threshold based step size control, and a Hampel stage is subsequently applied on the host PC for residual spike suppression. We characterize alpha-band modulation during stimulation and evaluate MAC in four healthy participants during head nodding, electrode tapping, and jaw clenching. The results support the feasibility of DRL stimulus based relative ESI mismatch monitoring for MAC in Ear-EEG devices.

\section{System Overview and Methodology}

\subsection{Overall System Architecture}

\begin{figure*}[!t]
    \centering
    \input{architecture_tikz}
    \caption{\textbf{Overview of the proposed system architecture.} (A) Placement of the DRL, reference electrode, and superiorly positioned left in-ear electrode (ELE) $Z_i$, (B) ELE electrode embedded within the earpiece, (C) ESI interface~\cite{chi2010}, (D) AFE with input buffers and fully differential instrumentation amplifiers, (E) DRL loop with stimulus injection, (F) stimulus signal scaling for safe body injection, (G) ESI extraction stage with filtering and multiplication-based I/Q demodulation, (H) anti-aliasing filters, (I) analog-to-digital converter (ADC), (J) microcontroller unit (MCU) with adaptive filtering, (K) Bluetooth Low Energy (BLE) module for wireless transmission, (L) reference signal estimation, and (M) stimulus generation with in-phase and quadrature phase signals.}
    \label{fig:overall_system}
    \vspace{-0.4cm}
\end{figure*}
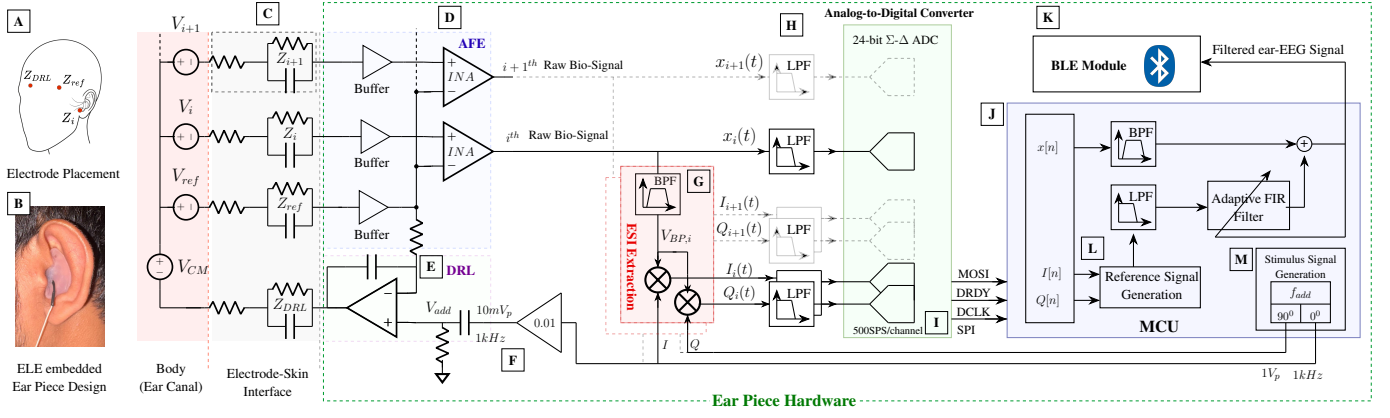

Fig.~\ref{fig:overall_system} shows the overall architecture of the proposed system, which integrates both biosignal acquisition and ESI reference


\noindent signal measurement. The setup uses one in-ear sensing electrode referenced to an external forehead electrode placed at Fp1, together with a DRL electrode. The same architecture can be extended to multichannel bipolar EEG acquisition.

The DRL feeds back the inverted common-mode estimate from the channels. It also injects a 1 kHz sinusoidal signal, \(V_{add}\). This signal appears in the common-mode voltage \(V_{cm}\) at the input of the differential amplifier~\cite{degen2007improved}. The injected frequency is selected so that it does not overlap with the EEG frequency range of interest (0.5-40~Hz). Here, the common-mode and differential-mode outputs of the differential amplifier are given by:
\begin{equation}
\small
V_{out} =
\underbrace{%
  \colorbox{purple!0}{$\displaystyle G_{DM}\,(V_{cm})\left(\frac{Z_i - Z_{ref}}{Z_{in}}\right)$}%
}_{\textcolor{black}{\text{Common Mode Signal}}}
\;+\;
\underbrace{%
  \colorbox{green!0}{$\displaystyle G_{DM}\,(V_i - V_{ref})$}%
}_{\textcolor{black}{\text{Differential Mode Signal}}}
\end{equation}

The combined output contains both the EEG signal and the ESI related signal, whereas the differential output contains the EEG signal. However when there is a mismatch between the electrode-skin impedance interfaces, i.e., \(Z_i - Z_{ref} \neq 0\), part of the common-mode signal appears together with the EEG signal~\cite{degen2007improved}. Since the EEG and ESI signals occupy different frequency bands, a 0-100~Hz lowpass filter (LPF) is used to extract the EEG signal, while a narrow bandpass filter (BPF) of 900-1200~Hz is used to extract the ESI signal.

\subsection{Analog Multiplication and I/Q Demodulation}

The extracted ESI signal ($V_{BP}$) was synchronously multiplied by the in-phase ($I$) and quadrature-phase ($Q$) stimulus signals and then low-pass filtered to remove double-frequency components. The resulting baseband components were sampled and used to estimate the amplitude and phase changes associated with the ESI mismatch $(Z_i - Z_{\mathrm{ref}})$ due to motion artifacts~\cite{degen2008continuous,cheon2024}. This extracted in-phase component is used as a relative ESI-mismatch-correlated reference rather than an absolute impedance estimate in ohms to evaluate the impact of motion artifacts on the Ear-EEG signal.

\begin{figure}[!b]
    \vspace{-0.6cm} 
    \centering
    \includegraphics[width=0.95\columnwidth]{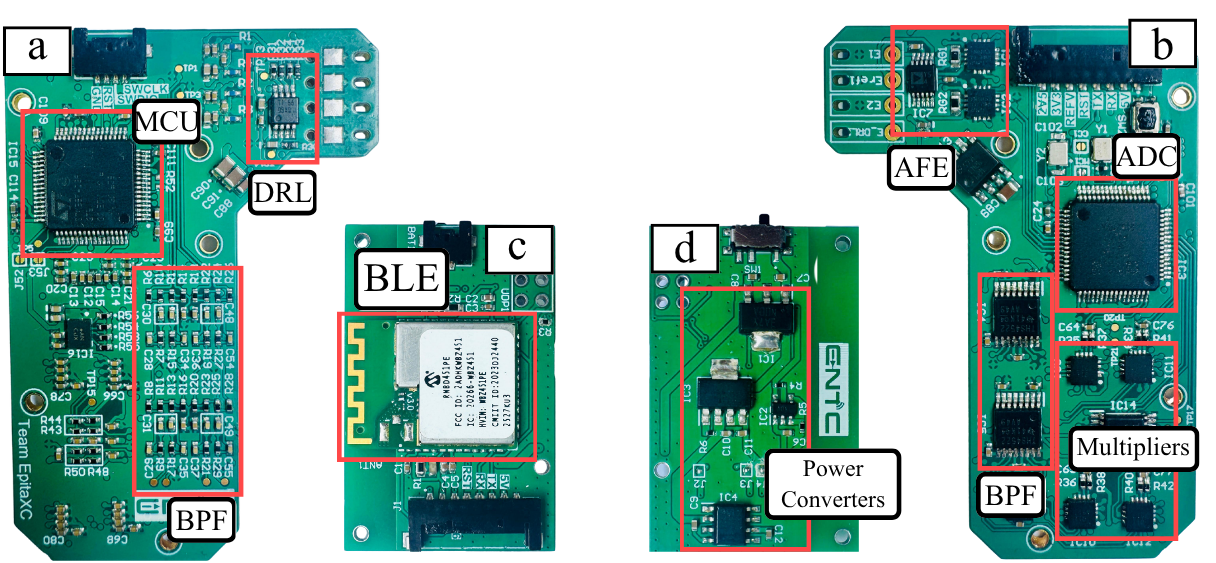}
    \caption{\textbf{Hardware implementation of the proposed system.} (a) Top-view of the acquisition circuit board. (b) Bottom-view of the acquisition circuit board. (c) BLE module for signal transmission and (d) power converters providing 5~V, 2.5~V, 3.3~V, and $-5$~V supply levels.}
    \label{fig:hardware_implementation}
\end{figure}

\subsection{Hardware Implementation}

Fig.~\ref{fig:hardware_implementation} shows the implementation of the system using commercially available off-the-shelf components. The AFE consists of input buffers, fully differential instrumentation amplifiers (INAs), and a DRL circuit. The design also includes a fourth-order fully differential bandpass filter for ESI extraction and fully differential multipliers for I/Q demodulation. Finally, both the Ear-EEG signals and the demodulated ESI signals are sampled using a 24-bit ADC and processed online on an STM32\textsuperscript{TM} microcontroller.

\subsection{Online Processing Algorithm Implementation}

A BLE module is used to wirelessly transmit the processed data to the graphical user interface with the real-time signal processing architecture. The real-time processing pipeline was designed to suppress motion-induced artifacts while preserving the EEG signal by taking into consideration that Rosanne {\textit{et al.}}\cite{10.3389/fnins.2021.611962} had shown that motion artifacts may span 0.11-20 Hz, overlapping with EEG bands of interest.

The raw EEG was bandpass filtered (0.5-40~Hz), and the impedance signal was low-pass filtered at 10~Hz. Both filters were implemented as 4th-order Butterworth infinite impulse response (IIR) filters. The ESI reference signal was low-pass filtered to capture slow variations in electrode-skin impedance mismatch caused by motion artifacts~\cite{muller}. The filtered signal was then used as the reference input to an adaptive finite impulse response (FIR) filter, while the filtered EEG signal served as the primary input.

The adaptive filter coefficients were updated using the NLMS rule\cite{inbook}, because the amplitude of the impedance reference can vary significantly during electrode-contact changes. The 64-tap NLMS filter used a base step size of $\mu_0=0.06$ and $\epsilon=10^{-3}$. In order to avoid over-adaptation during stable EEG periods and increase MAC only when electrode-contact variation was detected, the base NLMS step size was updated using a threshold-based approach. The step size was scaled by factors of 2, 3, 10, and 15 when the peak-to-peak ESI-reference variation, evaluated over a 64-sample window, exceeded 0.004, 0.008, 0.010, and 0.040~V, respectively. Adaptation was frozen when the variation exceeded 0.08~V and resumed when it fell to the normal range. These parameters were heuristically tuned using pilot recordings. A post-NLMS Hampel filter was used as in\cite{ghaleb2018two} with a local window of \(0.5\) s and a threshold of 3.0 scaled median absolute deviations.


\subsection{Data Collection Protocol}


The study was approved by the University of Moratuwa Ethics Review Committee under Ethics Declaration No. ERN/2026/003, with written consent obtained from all participants. Ear-EEG data were collected from four healthy participants, including one female and three males aged \(25 \pm 1\) years, using a custom-designed earpiece with an Ag/AgCl sintered pellet electrode. The DRL and reference electrodes were placed at \(\mathrm{Fpz}\) and \(\mathrm{Fp1}\), respectively, as shown in Fig.~\ref{fig:overall_system}.

\begin{figure}[!b]
    \vspace{-0.4cm}
    \centering
    \includegraphics[width=1\linewidth]{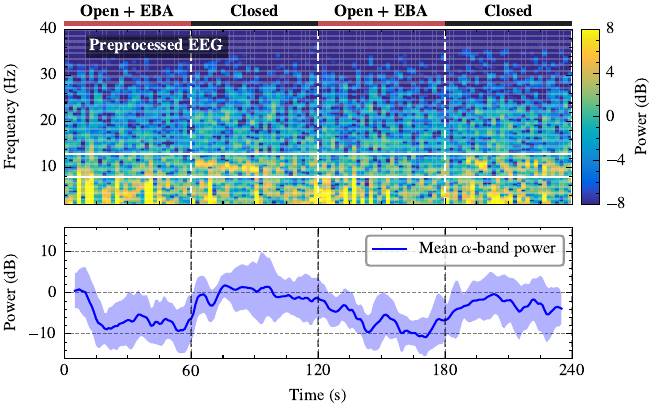}
    \vspace{-0.6cm}
    \caption{Time-frequency spectrogram of alpha modulation (top) using 4~s segments with 3~s overlap and grand average alpha power plot (bottom) for 12 recordings. EBA - Eye Blink Artifacts.}
    \label{fig:alpha}
\end{figure}


A 240~s alpha-wave protocol following~\cite{kappel2017physiological} was first used to characterize Ear-EEG acquisition and alpha-band (8-12~Hz) modulation after processing, as shown in Fig.~3. The 300~s artifact trials then evaluated head nodding, earpiece tapping, and jaw clenching. As shown in Fig.~4, each trial included five segments: eyes open without artifact from 0-50~s, eyes closed without artifact from 50-100~s, eyes closed with intentional artifact from 100-200~s, eyes closed without artifact from 200-250~s, and eyes open without artifact from 250-300~s.

\section{Results and Discussion}

The experimental validation focused on evaluating the effect of sinusoidal impedance stimulation on Ear-EEG quality, alpha wave modulation performance, and assessing the proposed impedance-aware digital signal processing pipeline for MAC.

\begin{figure*}[!t]
    \centering
    \begin{minipage}[t]{0.48\textwidth}
        \centering
        \includegraphics[width=\linewidth]{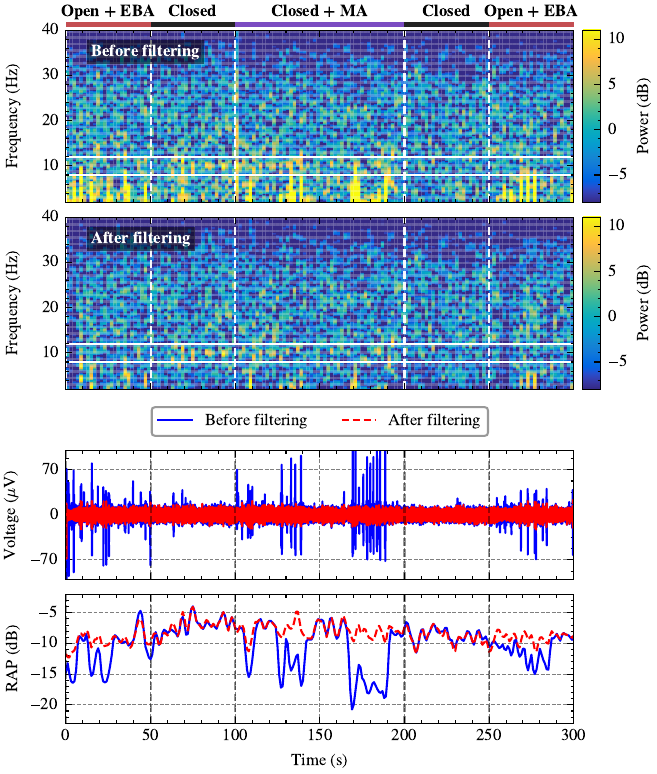}
        \vspace{-0.5cm}
        \centerline{\small (a) Electrode tapping condition}
    \end{minipage}
    \hfill
    \begin{minipage}[t]{0.48\textwidth}
        \centering
        \includegraphics[width=\linewidth]{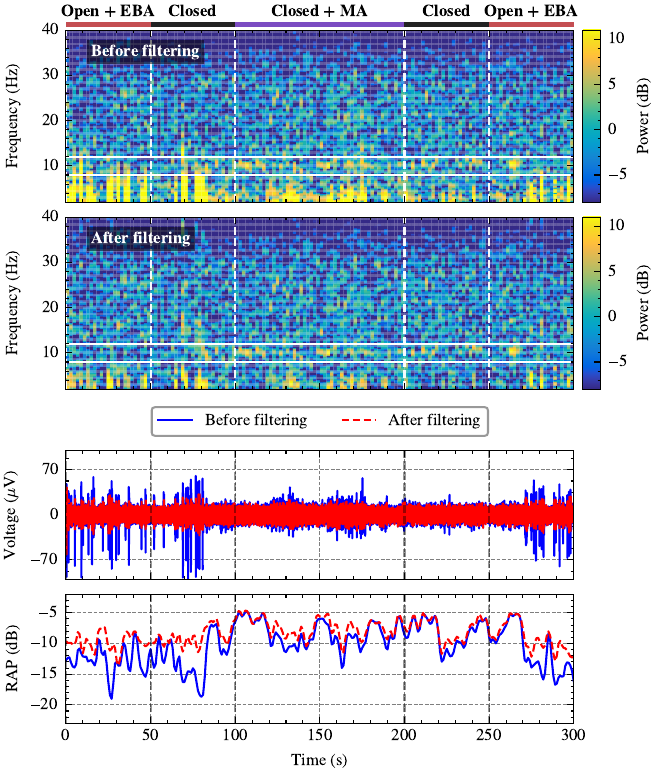}
        \vspace{-0.5cm}
        \centerline{\small (b) Head nodding condition}
    \end{minipage}
    \vspace{3mm}
    \caption{\textbf{Evaluation of the proposed impedance-aware motion artifact reduction method under two motion conditions.}
    (a) Measurement results for the electrode tapping condition. 
    (b) Measurement results for the head nodding condition. 
    Each result includes spectrograms before and after filtering, the time-domain comparison before and after filtering, and the relative alpha-band power comparison across experimental segments for two recordings from subject 2. The dashed vertical lines indicate the experimental segment boundaries, while the white horizontal lines in the spectrograms mark the alpha band. MA - Motion Artifacts.}
    \label{fig:motion_condition_comparison}
    \vspace{-0.4cm}
\end{figure*}

\subsection{Effect of Sinusoidal Stimulation}

The effect of sinusoidal DRL stimulation on Ear-EEG quality was evaluated by comparing eyes-closed recordings with and without 1~kHz stimulation, where alpha-band activity is expected to be prominent. Effect on alpha-band was quantified using relative alpha power ($\mathrm{RAP}_{\mathrm{dB}}$), as described in~\cite{10.3389/fnins.2024.1441897}:
\begin{equation}
\mathrm{RAP}_{\mathrm{dB}} =
10\log_{10}
 \left(
\frac{
P_{8\text{ to }12}
}
{
P_{0.5\text{ to }40} - P_{7\text{ to }13}
}
 \right)
\label{eq:rap_stimulation}
\end{equation}

where \(P_{8\text{ to }12}\) is the alpha-band power, \(P_{0.5\text{ to }40}\) is the total EEG power, and \(P_{7\text{ to }13}\) is the excluded alpha-surrounding band. The average RAP values were -2.15~dB without stimulation and -2.21~dB with stimulation, corresponding to only a 0.06~dB difference. This indicates that the 1~kHz DRL stimulus supported impedance monitoring without noticeably degrading alpha-band activity.

\begin{table}[t!]
\caption{Individual and Grand Average Alpha Modulation Ratios}
\begin{center}
\renewcommand{\arraystretch}{1.2}
\begin{tabular}{p{2.2cm}p{2.4cm}}
\hline
 & \textbf{Ear EEG ($\text{V}^2/\text{V}^2$)} \\
\hline
Subject 1 & $1.97 \pm 0.89$ \\
Subject 2 & $2.03 \pm 0.08$ \\
Subject 3 & $1.84 \pm 0.80$ \\
Subject 4 & $1.29 \pm 0.28$ \\
\hline
\textbf{Average} & $\mathbf{1.78 \pm 0.33}$ \\
\hline
\end{tabular}
\label{tab:alpha_modulation_ratios}
\end{center}
\vspace{-0.6cm}
\end{table}


\subsection{Alpha Wave Modulation}

The alpha modulation ratio (AMR) was calculated as follows~\cite{Kaveh:EECS-2023-56}:
\vspace{-0.1cm}
\begin{equation}
\mathrm{AMR} =
\frac{
\text{mean alpha band power (eyes closed)}
}{
\text{mean alpha band power (eyes opened)}
}
\label{eq}
\end{equation}

As shown in Fig.~\ref{fig:alpha} and Table~\ref{tab:alpha_modulation_ratios}, the recordings achieved a grand average ratio of \(1.78 \pm 0.33\), confirming detectable alpha modulation across the four participants. This was slightly lower than the reported \(2.17 \pm 0.69\)~\cite{Kaveh:EECS-2023-56}, likely due to dry electrode contact quality and the limited cohort size.

\subsection{Motion Artifact Cancellation}

MAC was evaluated for head nodding, electrode tapping, and jaw clenching using 12 recordings per condition from four participants. Artifact reduction was assessed using spectrogram inspection, average artifact band power reduction, and relative alpha band power improvement as shown in
Table~\ref{tab:motion_artifact_reduction}. The power spectral density (PSD) of the Ear-EEG signal was estimated using Welch's method\cite{1161901} with a segment size of 4~s and an overlap between segments of 3~s, and the artifact power was represented by the mean PSD within the 2 to 7 Hz artifact-dominated band. For artifacts, this metric quantifies only the low-frequency contact-related component, not the broadband myogenic EMG contamination\cite{kappel2017physiological}.

The artifact power reduction in decibels ($R_{\mathrm{dB}}$) and the corresponding percentage artifact power reduction ($R_{\%}$) was calculated using the mean PSD values. $R_{\mathrm{dB}}$ and $R_{\%}$ were computed separately for each recording, averaged within each participant, and then averaged across the four participant-level values.
\vspace{-0.05cm}
\begin{equation}
R_{\mathrm{dB}} =
P_{2\text{--}7,\mathrm{before,dB}} -
P_{2\text{--}7,\mathrm{after,dB}}
\label{eq:artifact_reduction_db}
\end{equation}

\vspace{-0.1 cm}

\begin{equation}
R_{\%} =
\left(
\frac{
P_{2\text{--}7,\mathrm{before}} -
P_{2\text{--}7,\mathrm{after}}
}
{
P_{2\text{--}7,\mathrm{before}}
}
\right)
\times 100\%
\label{eq:percentage_reduction}
\end{equation}


\begin{table}[!b]
\vspace{-0.6cm}
\caption{Motion Artifact Cancellation Results}
\begin{center}
\renewcommand{\arraystretch}{1.2}
\begin{tabular}{p{1.9cm}p{1.7cm}p{1.7cm}p{1.2cm}}
\hline
\textbf{Motion \hspace{1cm} condition} & 
$\boldsymbol{R_{\mathrm{dB}}}$ & 
$\boldsymbol{R_{\%}}$ & 
\textbf{RAP change} \\
\hline
Head nodding & 6.52~dB & 77.6\% & 2.3~dB \\
Tapping & 12.56~dB & 92.8\% & 6~dB \\
Jaw clenching & 9.03~dB & 86.4\% & 2.5~dB \\
\hline
\end{tabular}
\label{tab:motion_artifact_reduction}
\end{center}
\end{table}

RAP was computed from the mean PSD in the 8-12~Hz alpha band relative to the surrounding non-alpha EEG content, with RAP change defined as the difference between the values after and before the motion artifact cancellation pipeline, as summarized in Table~\ref{tab:motion_artifact_reduction}. Fig.~\ref{fig:motion_condition_comparison} shows reduced low-frequency artifacts and time-domain variations while preserving observable alpha-band modulation. The Hampel stage replaced $<1\%$ of samples on average, indicating only sparse residual spike correction. The strongest artifact reduction occurred during electrode tapping, likely because it directly affects the electrode-skin interface and is well captured by the ESI reference, whereas head nodding and jaw clenching also introduce EMG components not directly represented by this reference. Overall, the results support effective impedance-referenced artifact attenuation, although ground-truth validation is still needed to rigorously assess broadband neural-signal preservation.

\section{Conclusion and Future Work}

This study demonstrates the feasibility of a wireless dry contact Ear-EEG system using a DRL based sinusoidal stimulus for continuous relative electrode-skin impedance mismatch monitoring. The derived impedance variations served as an adaptive reference for motion artifact cancellation, achieving artifact power reductions of 77.6\%, 92.8\%, and 86.4\% during head nodding, electrode tapping, and jaw clenching, respectively, while alpha band modulation remained observable. These results support impedance aware processing for wearable Ear-EEG during motion. Future work will include larger cohorts, ground truth validation, phantom based electrical validation, integration of a more compact earpiece with a fully in-ear electrode configuration, and extension to motion artifact suppression in other wearable biosensing modalities.




\color{black}

\bibliographystyle{IEEEtran}
\bibliography{ref}

\end{document}

%% file: architecture_tikz.tex
\resizebox{1\textwidth}{!}{%
    \begin{circuitikz}
    \tikzstyle{every node}=[font=\fontsize{13.1pt}{17.0pt}\selectfont]
    \draw [ color={rgb,255:red,0; green,51; blue,255}, draw opacity=0.16 , fill={rgb,255:red,168; green,183; blue,255}, fill opacity=0.05, line width=1pt , dashed] (-6.25,11) rectangle  (-0.625,3.75);
    \draw [ color={rgb,255:red,20; green,30; blue,140}, draw opacity=0.54 , fill={rgb,255:red,186; green,191; blue,255}, fill opacity=0.14, line width=1pt ] (16.75,8.625) rectangle (28.625,0.75);
    \node [font=\fontsize{13.1pt}{17.0pt}\selectfont, fill={rgb,255:red,255; green,255; blue,255}, fill opacity=0, text opacity=1, inner xsep=0.080cm, inner ysep=0.085cm, rounded corners=0.020cm] at (26.625,2.25) {$f_{add}$};
    \node [font=\fontsize{10.8pt}{14.1pt}\selectfont, fill={rgb,255:red,255; green,255; blue,255}, fill opacity=0, text opacity=1, inner xsep=0.080cm, inner ysep=0.085cm, rounded corners=0.020cm, align=center] at (26.625,3.125) {Stimulus Signal \\ Generation};
    \draw [ line width=1pt ] (25.625,2.625) rectangle (27.625,1.25);
    \draw [ line width=1pt ] (25.125,3.75) rectangle (28.125,1);
    \draw [ color={rgb,255:red,7; green,158; blue,55}, draw opacity=1 , line width=1pt , dashed] (-6.25,12) rectangle  (29,-1.375);
    \node [font=\fontsize{11.9pt}{15.5pt}\selectfont, fill={rgb,255:red,255; green,255; blue,255}, fill opacity=0, text opacity=1, inner xsep=0.080cm, inner ysep=0.085cm, rounded corners=0.020cm] at (5.25,6) {BPF};
    \draw [ line width=1pt ] (4.25,6.25) rectangle (5.75,4.875);
    \draw [ color={rgb,255:red,212; green,0; blue,0}, draw opacity=0.35 , fill={rgb,255:red,248; green,173; blue,173}, fill opacity=0.11, line width=1pt ] (3.75,6.5) rectangle (6.875,1.25);
    \node [font=\fontsize{14.8pt}{19.2pt}\selectfont, fill={rgb,255:red,255; green,255; blue,255}, fill opacity=0, text opacity=1, inner xsep=0.080cm, inner ysep=0.085cm, rounded corners=0.020cm] at (7.75,2.25) {$Q_i(t)$};
    \node [font=\fontsize{14.8pt}{19.2pt}\selectfont, fill={rgb,255:red,255; green,255; blue,255}, fill opacity=0, text opacity=1, inner xsep=0.080cm, inner ysep=0.085cm, rounded corners=0.020cm] at (7.75,3) {$I_i(t)$};
    \node [font=\fontsize{14.8pt}{19.2pt}\selectfont, fill={rgb,255:red,255; green,255; blue,255}, fill opacity=0, text opacity=1, inner xsep=0.080cm, inner ysep=0.085cm, rounded corners=0.020cm] at (7.75,4.375) {$Q_{i+1}(t)$};
    \node [font=\fontsize{14.8pt}{19.2pt}\selectfont, fill={rgb,255:red,255; green,255; blue,255}, fill opacity=0, text opacity=1, inner xsep=0.080cm, inner ysep=0.085cm, rounded corners=0.020cm] at (7.75,5.25) {$I_{i+1}(t)$};
    \draw [ color={rgb,255:red,87; green,153; blue,55}, draw opacity=0.47 , fill={rgb,255:red,186; green,255; blue,203}, fill opacity=0.12, line width=1pt ] (11.25,11.25) rectangle (14.875,0.75);
    \draw [ dashed] (-10,11) rectangle  (-6.5,9);
    \node [font=\fontsize{13.1pt}{17.0pt}\selectfont, fill={rgb,255:red,255; green,255; blue,255}, fill opacity=0, text opacity=1, inner xsep=0.080cm, inner ysep=0.085cm, rounded corners=0.020cm] at (25.625,-0.5) {$1V_p$};
    \node [font=\fontsize{13.1pt}{17.0pt}\selectfont, fill={rgb,255:red,255; green,255; blue,255}, fill opacity=0, text opacity=1, inner xsep=0.080cm, inner ysep=0.085cm, rounded corners=0.020cm] at (26.875,-0.5) {$1 kHz$};
    \node [font=\fontsize{13.1pt}{17.0pt}\selectfont, fill={rgb,255:red,255; green,255; blue,255}, fill opacity=0, text opacity=1, inner xsep=0.080cm, inner ysep=0.085cm, rounded corners=0.010cm] at (1.2,1.25) {\textbf{$0.01$}};
    \draw [ color={rgb,255:red,0; green,0; blue,0}, draw opacity=0 , fill={rgb,255:red,244; green,244; blue,244}, fill opacity=0.5, line width=1pt ] (-10,11) rectangle (-6.375,0.625);
    \node [font=\fontsize{14.2pt}{18.5pt}\selectfont, fill={rgb,255:red,255; green,255; blue,255}, fill opacity=0, text opacity=1, inner xsep=0.080cm, inner ysep=0.085cm, rounded corners=0.020cm] at (-7.375,1.86) {$Z_{DRL}$};
    \node [font=\fontsize{14.2pt}{18.5pt}\selectfont, fill={rgb,255:red,255; green,255; blue,255}, fill opacity=0, text opacity=1, inner xsep=0.080cm, inner ysep=0.085cm, rounded corners=0.020cm] at (-7.375,5.215) {$Z_{ref}$};
    \node [font=\fontsize{14.2pt}{18.5pt}\selectfont, fill={rgb,255:red,255; green,255; blue,255}, fill opacity=0, text opacity=1, inner xsep=0.080cm, inner ysep=0.085cm, rounded corners=0.020cm] at (-7.375,7.59) {$Z_{i}$};
    \node [font=\fontsize{14.2pt}{18.5pt}\selectfont, fill={rgb,255:red,255; green,255; blue,255}, fill opacity=0, text opacity=1, inner xsep=0.080cm, inner ysep=0.085cm, rounded corners=0.020cm] at (-7.375,10.09) {$Z_{i+1}$};
    \node [font=\fontsize{10.2pt}{13.3pt}\selectfont, fill={rgb,255:red,255; green,255; blue,255}, fill opacity=0, text opacity=1, inner xsep=0.080cm, inner ysep=0.085cm, rounded corners=0.020cm] at (12.75,1) {500SPS/channel};
    \node [font=\fontsize{13.1pt}{17.0pt}\selectfont, fill={rgb,255:red,255; green,255; blue,255}, fill opacity=0, text opacity=1, inner xsep=0.080cm, inner ysep=0.085cm, rounded corners=0.020cm] at (9.75,9.875) {LPF};
    \draw [ line width=1pt ] (20.25,5.875) rectangle (21.75,4.375);
    \draw [ line width=1pt ] (17.375,8.25) rectangle (19,1.25);
    \node [font=\fontsize{11.9pt}{15.5pt}\selectfont, fill={rgb,255:red,255; green,255; blue,255}, fill opacity=0, text opacity=1, inner xsep=0.080cm, inner ysep=0.085cm, rounded corners=0.020cm] at (15.625,1.625) {DCLK};
    \node [font=\fontsize{11.9pt}{15.5pt}\selectfont, fill={rgb,255:red,255; green,255; blue,255}, fill opacity=0, text opacity=1, inner xsep=0.080cm, inner ysep=0.085cm, rounded corners=0.020cm] at (15.625,2.25) {DRDY};
    \node [font=\fontsize{11.9pt}{15.5pt}\selectfont, fill={rgb,255:red,255; green,255; blue,255}, fill opacity=0, text opacity=1, inner xsep=0.080cm, inner ysep=0.085cm, rounded corners=0.020cm] at (15.625,2.875) {MOSI};
    \node [font=\fontsize{13.1pt}{17.0pt}\selectfont, fill={rgb,255:red,255; green,255; blue,255}, fill opacity=0, text opacity=1, inner xsep=0.080cm, inner ysep=0.085cm, rounded corners=0.020cm] at (25.875,10.375) {Filtered ear-EEG Signal};
    \draw [ line width=1pt ] (17.625,10.875) rectangle (23.25,9);
    \node [font=\fontsize{13.1pt}{17.0pt}\selectfont, fill={rgb,255:red,255; green,255; blue,255}, fill opacity=0, text opacity=1, inner xsep=0.080cm, inner ysep=0.085cm, rounded corners=0.010cm] at (26.75,7.25) {+};
    \draw [ line width=1pt ] (26.75,7.25) circle (0.25cm);
    \node [font=\fontsize{13.1pt}{17.0pt}\selectfont, fill={rgb,255:red,255; green,255; blue,255}, fill opacity=0, text opacity=1, inner xsep=0.080cm, inner ysep=0.085cm, rounded corners=0.020cm, align=center] at (21.625,2.5) {Reference Signal \\ Generation};
    \draw [ color={rgb,255:red,212; green,0; blue,0}, draw opacity=0.35 , fill={rgb,255:red,255; green,168; blue,168}, fill opacity=0.03, line width=1pt , dashed] (3.75,6.5) rectangle  (6.875,1.25);
    \node [font=\fontsize{13.1pt}{17.0pt}\selectfont, color={rgb,255:red,231; green,0; blue,0}, text opacity=1, , rotate around={-90:(0,0)}, fill={rgb,255:red,255; green,255; blue,255}, fill opacity=0, text opacity=1, inner xsep=0.080cm, inner ysep=0.085cm, rounded corners=0.020cm] at (4.125,3.125) {\textbf{ESI Extraction}};
    \node [font=\fontsize{13.1pt}{17.0pt}\selectfont, fill={rgb,255:red,255; green,255; blue,255}, fill opacity=0, text opacity=1, inner xsep=0.080cm, inner ysep=0.085cm, rounded corners=0.020cm] at (21.25,5.5) {LPF};
    \draw [ color={rgb,255:red,161; green,1; blue,177}, draw opacity=0.14 , fill={rgb,255:red,180; green,255; blue,168}, fill opacity=0.03, line width=1pt , dashed] (-6.25,3.5) rectangle  (-0.625,0.625);
    \draw [line width=1pt, -{Stealth[scale=1.5]}, ] (23.75,4.125) -- (25.625,6.375);
    \draw [ color={rgb,255:red,0; green,0; blue,0}, draw opacity=0 , fill={rgb,255:red,255; green,222; blue,222}, fill opacity=0.5, line width=1pt ] (-12.5,11) rectangle (-10.125,0.625);
    \node [font=\fontsize{15.9pt}{20.7pt}\selectfont, fill={rgb,255:red,255; green,255; blue,255}, fill opacity=0, text opacity=1, inner xsep=0.080cm, inner ysep=0.085cm, rounded corners=0.020cm] at (-10.875,11.25) {$V_{i+1}$};
    \node [font=\fontsize{13.1pt}{17.0pt}\selectfont, fill={rgb,255:red,255; green,255; blue,255}, fill opacity=0, text opacity=1, inner xsep=0.080cm, inner ysep=0.085cm, rounded corners=0.020cm, align=center] at (-11.375,-0.625) {Body \\ (Ear Canal)};
    \node [font=\fontsize{13.1pt}{17.0pt}\selectfont, fill={rgb,255:red,255; green,255; blue,255}, fill opacity=0, text opacity=1, inner xsep=0.080cm, inner ysep=0.085cm, rounded corners=0.020cm] at (-4.625,4.225) {Buffer};
    \node [font=\fontsize{13.1pt}{17.0pt}\selectfont, fill={rgb,255:red,255; green,255; blue,255}, fill opacity=0, text opacity=1, inner xsep=0.080cm, inner ysep=0.085cm, rounded corners=0.020cm] at (-4.625,6.625) {Buffer};
    \node [font=\fontsize{13.1pt}{17.0pt}\selectfont, fill={rgb,255:red,255; green,255; blue,255}, fill opacity=0, text opacity=1, inner xsep=0.080cm, inner ysep=0.085cm, rounded corners=0.020cm] at (-4.625,9.125) {Buffer};
    \node [font=\fontsize{13.1pt}{17.0pt}\selectfont, fill={rgb,255:red,255; green,255; blue,255}, fill opacity=0, text opacity=1, inner xsep=0.080cm, inner ysep=0.085cm, rounded corners=0.020cm] at (-1.625,7) {$INA$};
    \node [font=\fontsize{13.1pt}{17.0pt}\selectfont, fill={rgb,255:red,255; green,255; blue,255}, fill opacity=0, text opacity=1, inner xsep=0.080cm, inner ysep=0.085cm, rounded corners=0.020cm] at (-1.625,9.5) {$INA$};
    \draw [ line width=0.8pt](-1.375,9.5) node[op amp,scale=1, yscale=-1 ] (opamp2) {};
    \draw [ line width=0.8pt](opamp2.+) to[short] (-2.875,10);
    \draw [ line width=0.8pt] (opamp2.-) to[short] (-2.875,9);
    \draw [ line width=0.8pt](-0.175,9.5) to[short](0.125,9.5);
    \draw [ line width=0.8pt](-1.375,7) node[op amp,scale=1, yscale=-1 ] (opamp2) {};
    \draw [ line width=0.8pt](opamp2.+) to[short] (-2.875,7.5);
    \draw [ line width=0.8pt] (opamp2.-) to[short] (-2.875,6.5);
    \draw [ line width=0.8pt](-0.175,7) to[short](0.125,7);
    \draw (-5.3125,10) node[ieeestd buffer port, anchor=in](port){} (port.out) to[short] (-3.5,10);
    \draw (port.in) to[short] (-5.625,10);
    \draw (-5.375,7.5) node[ieeestd buffer port, anchor=in](port){} (port.out) to[short] (-3.625,7.5);
    \draw (port.in) to[short] (-5.625,7.5);
    \draw (-5.4375,5.125) node[ieeestd buffer port, anchor=in](port){} (port.out) to[short] (-3.625,5.125);
    \draw (port.in) to[short] (-5.75,5.125);
    \draw [ line width=0.8pt](-3.125,5.125) to[R] (-3.125,3.125);
    \draw [line width=0.8pt](-3.75,3.125) to[C] (-5.5,3.125);
    \draw [ line width=1pt](-2.625,9) to[short] (-3.125,9);
    \draw [ line width=1pt](-3.125,5.125) to[short] (-3.125,9);
    \draw [ line width=1pt](-2.875,7.5) to[short] (-3.625,7.5);
    \draw [ line width=1pt](-3.625,5.125) to[short] (-3.125,5.125);
    \draw [ line width=1pt](-3.125,6.5) to[short] (-2.625,6.5);
    \draw [ line width=1pt](-3.5,10) to[short] (-2.875,10);
    \draw [ line width=0.8pt](-10.25,10) to[R] (-8.75,10);
    \draw [ line width=0.8pt](-8.125,10.625) to[R] (-6.625,10.625);
    \draw [line width=0.8pt](-8.125,9.375) to[C] (-6.625,9.375);
    \draw [ line width=1pt](-8.125,10.625) to[short] (-8.125,9.375);
    \draw [ line width=1pt](-8.75,10) to[short] (-8.125,10);
    \draw [ line width=1pt](-6.625,10.625) to[short] (-6.625,9.375);
    \draw [ line width=1pt](-5.625,10) to[short] (-6.625,10);
    \draw [ line width=0.8pt](-10.25,7.5) to[R] (-8.75,7.5);
    \draw [ line width=0.8pt](-8.125,8.125) to[R] (-6.625,8.125);
    \draw [line width=0.8pt](-8.125,6.875) to[C] (-6.625,6.875);
    \draw [ line width=1pt](-8.125,8.125) to[short] (-8.125,6.875);
    \draw [ line width=1pt](-8.75,7.5) to[short] (-8.125,7.5);
    \draw [ line width=1pt](-6.625,8.125) to[short] (-6.625,6.875);
    \draw [ line width=1pt](-5.625,7.5) to[short] (-6.625,7.5);
    \draw [ line width=0.8pt](-10.25,5.125) to[R] (-8.75,5.125);
    \draw [ line width=0.8pt](-8.125,5.75) to[R] (-6.625,5.75);
    \draw [line width=0.8pt](-8.125,4.5) to[C] (-6.625,4.5);
    \draw [ line width=1pt](-8.125,5.75) to[short] (-8.125,4.5);
    \draw [ line width=1pt](-8.75,5.125) to[short] (-8.125,5.125);
    \draw [ line width=1pt](-6.625,5.75) to[short] (-6.625,4.5);
    \draw [ line width=1pt](-5.625,5.125) to[short] (-6.625,5.125);
    \draw [ line width=1pt](-3.125,3.125) to[short] (-3.125,2.25);
    \draw [ line width=1pt](-3.75,3.125) to[short] (-3.125,3.125);
    \draw [ line width=1pt](-6.125,1.75) to[short] (-6.125,3.125);
    \draw [ line width=1pt](-5.5,3.125) to[short] (-6.125,3.125);
    \draw [ line width=1pt](-6.125,1.75) to[short] (-6.625,1.75);
    \draw [ line width=0.8pt](-10.25,1.75) to[R] (-8.75,1.75);
    \draw [ line width=0.8pt](-8.125,2.375) to[R] (-6.625,2.375);
    \draw [line width=0.8pt](-8.125,1.125) to[C] (-6.625,1.125);
    \draw [ line width=1pt](-8.125,2.375) to[short] (-8.125,1.125);
    \draw [ line width=1pt](-8.75,1.75) to[short] (-8.125,1.75);
    \draw [ line width=1pt](-6.625,2.375) to[short] (-6.625,1.125);
    \draw [ line width=0.8pt](-4.625,1.75) node[op amp,scale=1, xscale=-1 ] (opamp2) {};
    \draw [ line width=0.8pt](opamp2.+) to[short] (-3.125,1.25);
    \draw [ line width=0.8pt] (opamp2.-) to[short] (-3.125,2.25);
    \draw [ line width=0.8pt](-5.825,1.75) to[short](-6.125,1.75);
    \draw [ line width=0.8pt](-11.5,5.125) to[american voltage source] (-10.25,5.125);
    \draw [ line width=0.8pt](-11.5,7.5) to[american voltage source] (-10.25,7.5);
    \draw [ line width=0.8pt](-11.5,10) to[american voltage source] (-10.25,10);
    \draw [ line width=0.8pt](-11.75,3.875) to[american voltage source] (-11.75,2.375);
    \draw [ line width=1pt](-11.75,2.375) to[short] (-11.75,1.75);
    \draw [ line width=1pt](-10.375,1.75) to[short] (-11.75,1.75);
    \draw [ line width=1pt](-11.75,3.875) to[short] (-11.75,10);
    \draw [ line width=1pt](-11.75,10) to[short] (-11.25,10);
    \draw [ line width=1pt](-11.25,7.5) to[short] (-11.75,7.5);
    \draw [ line width=1pt](-11.25,5.125) to[short] (-11.75,5.125);
    \draw [ line width=1pt](-10.25,1.75) to[short] (-10.875,1.75);
    \draw [line width=1pt, -{Stealth[scale=1.5]}, ] (4.5,5.125) -- (5.625,5.125);
    \draw [line width=1pt, short] (4.625,5.125) -- (4.75,5.75);
    \draw [line width=1pt, short] (4.75,5.75) -- (5.25,5.75);
    \draw [line width=1pt, short] (5.25,5.75) -- (5.375,5.125);
    \node [font=\fontsize{13.1pt}{17.0pt}\selectfont, fill={rgb,255:red,255; green,255; blue,255}, fill opacity=0, text opacity=1, inner xsep=0.080cm, inner ysep=0.085cm, rounded corners=0.000cm] at (5.25,0.5) {$I$};
    \node [font=\fontsize{13.1pt}{17.0pt}\selectfont, fill={rgb,255:red,255; green,255; blue,255}, fill opacity=0, text opacity=1, inner xsep=0.080cm, inner ysep=0.085cm, rounded corners=0.020cm] at (6.25,0.5) {$Q$};
    \node [font=\fontsize{10.8pt}{14.1pt}\selectfont, fill={rgb,255:red,255; green,255; blue,255}, fill opacity=0, text opacity=1, inner xsep=0.080cm, inner ysep=0.085cm, rounded corners=0.020cm] at (26.125,1.5) {$90^0$};
    \node [font=\fontsize{10.8pt}{14.1pt}\selectfont, fill={rgb,255:red,255; green,255; blue,255}, fill opacity=0, text opacity=1, inner xsep=0.080cm, inner ysep=0.085cm, rounded corners=0.020cm] at (27.125,1.5) {$0^0$};
    \draw [line width=1pt, short] (26.625,1.875) -- (26.625,1.25);
    \draw [line width=1pt, short] (25.625,1.875) -- (27.625,1.875);
    \draw [ line width=1pt](0.125,7) to[short] (8.75,7);
    \draw [ line width=1pt](5,7) to[short] (5,6.25);
    \draw [ color={rgb,255:red,0; green,0; blue,0}, draw opacity=0.3, line width=1pt, dashed] (0.25,9.5) -- (8.75,9.5);
    \draw [ color={rgb,255:red,0; green,0; blue,0}, draw opacity=0.27, line width=1pt, dashed] (3.5,9.5) -- (3.5,6.125);
    \draw (-4.625,1.75) node[op amp,scale=1, xscale=-1 ] (opamp2) {};
    \draw (opamp2.+) to[short] (-3.125,1.25);
    \draw  (opamp2.-) to[short] (-3.125,2.25);
    \draw (-5.825,1.75) to[short](-6.125,1.75);
    \draw [line width=1pt, dashed] (-3.125,9) -- (-3.125,11.125);
    \draw [line width=1pt, dashed] (-11.75,10) -- (-11.75,11);
    \node [font=\fontsize{15.9pt}{20.7pt}\selectfont, fill={rgb,255:red,255; green,255; blue,255}, fill opacity=0, text opacity=1, inner xsep=0.080cm, inner ysep=0.085cm, rounded corners=0.020cm] at (-10.875,8.5) {$V_{i}$};
    \node [font=\fontsize{15.9pt}{20.7pt}\selectfont, fill={rgb,255:red,255; green,255; blue,255}, fill opacity=0, text opacity=1, inner xsep=0.080cm, inner ysep=0.085cm, rounded corners=0.020cm] at (-10.875,6.125) {$V_{ref}$};
    \node [font=\fontsize{13.1pt}{17.0pt}\selectfont, fill={rgb,255:red,255; green,255; blue,255}, fill opacity=0, text opacity=1, inner xsep=0.080cm, inner ysep=0.085cm, rounded corners=0.020cm] at (0.125,7.5) {$i^{th}$};
    \node [font=\fontsize{15.9pt}{20.7pt}\selectfont, fill={rgb,255:red,255; green,255; blue,255}, fill opacity=0, text opacity=1, inner xsep=0.080cm, inner ysep=0.085cm, rounded corners=0.020cm] at (7.75,7.5) {$x_i(t)$};
    \node [font=\fontsize{13.1pt}{17.0pt}\selectfont, fill={rgb,255:red,255; green,255; blue,255}, fill opacity=0, text opacity=1, inner xsep=0.080cm, inner ysep=0.085cm, rounded corners=0.020cm] at (-0.5,1.625) {$10mV_p$};
    \node [font=\fontsize{13.1pt}{17.0pt}\selectfont, color={rgb,255:red,34; green,1; blue,182}, text opacity=1, , fill={rgb,255:red,255; green,255; blue,255}, fill opacity=0, text opacity=1, inner xsep=0.080cm, inner ysep=0.085cm, rounded corners=0.020cm] at (-1.25,10.625) {\textbf{AFE}};
    \begin{scope}[transparency group, opacity=0]
    \node at (-3.125,6.5) [circ, color={rgb,255:red,0; green,0; blue,0}] {};
    \end{scope}
    \begin{scope}[transparency group, opacity=1]
    \node at (-3.125,6.5) [circ, color={rgb,255:red,0; green,0; blue,0}] {};
    \end{scope}
    \begin{scope}[transparency group, opacity=1]
    \node at (-3.125,5.125) [circ, color={rgb,255:red,0; green,0; blue,0}] {};
    \end{scope}
    \begin{scope}[transparency group, opacity=1]
    \node at (-3.125,9) [circ, color={rgb,255:red,0; green,0; blue,0}] {};
    \end{scope}
    \node [font=\fontsize{13.1pt}{17.0pt}\selectfont, color={rgb,255:red,116; green,0; blue,146}, text opacity=1, , fill={rgb,255:red,255; green,255; blue,255}, fill opacity=0, text opacity=1, inner xsep=0.080cm, inner ysep=0.085cm, rounded corners=0.020cm] at (-1.625,3) {\textbf{DRL}};
    \node [font=\fontsize{13.1pt}{17.0pt}\selectfont, fill={rgb,255:red,255; green,255; blue,255}, fill opacity=0, text opacity=1, inner xsep=0.080cm, inner ysep=0.085cm, rounded corners=0.020cm, align=center] at (-8.125,-0.75) {Electrode-Skin \\ Interface};
    \draw [ color={rgb,255:red,255; green,143; blue,124}, draw opacity=1, line width=1pt, dashed] (-10.125,11) -- (-10.125,-0.375);
    \draw [ color={rgb,255:red,0; green,0; blue,0}, draw opacity=0.31, line width=1pt, dashed] (-6.375,11) -- (-6.375,-0.375);
    \node [font=\fontsize{15.9pt}{20.7pt}\selectfont, color={rgb,255:red,0; green,122; blue,4}, text opacity=1, , fill={rgb,255:red,255; green,255; blue,255}, fill opacity=1, text opacity=1, inner xsep=0.080cm, inner ysep=0.085cm, rounded corners=0.020cm] at (9.25,-1.375) {\textbf{Ear Piece Hardware}};
    \draw [ line width=1pt](6,2.875) to[lamp] (6,1);
    \draw [ line width=1pt](5,3.5) to[lamp] (5,1.875);
    \draw [ line width=1pt](5,3.375) to[short] (5,4.875);
    \draw [ line width=1pt](5,3.625) to[short] (5.625,3.625);
    \draw [line width=1pt, short] (8.75,1.875) -- (6.375,1.875);
    \node [font=\fontsize{11.9pt}{15.5pt}\selectfont, fill={rgb,255:red,255; green,255; blue,255}, fill opacity=0, text opacity=1, inner xsep=0.080cm, inner ysep=0.085cm, rounded corners=0.020cm] at (13.125,11.625) {\textbf{Analog-to-Digital Converter}};
    \node [font=\fontsize{15.9pt}{20.7pt}\selectfont, fill={rgb,255:red,255; green,255; blue,255}, fill opacity=0, text opacity=1, inner xsep=0.080cm, inner ysep=0.085cm, rounded corners=0.020cm] at (7.75,10) {$x_{i+1}(t)$};
    \draw [ line width=1pt ] (19.875,3.125) rectangle (23.25,1.75);
    \draw [line width=1pt, -{Stealth[scale=1.5]}, ] (19,2) -- (19.875,2);
    \draw [line width=1pt, -{Stealth[scale=1.5]}, ] (19,2.875) -- (19.875,2.875);
    \draw [line width=1pt, -{Stealth[scale=1.5]}, ] (20.5,4.625) -- (20.5,5.625);
    \draw [line width=1pt, -{Stealth[scale=1.5]}, ] (20.5,4.625) -- (21.625,4.625);
    \draw [line width=1pt, short] (20.5,5.25) -- (21,5.25);
    \draw [line width=1pt, short] (21,5.25) -- (21.125,4.625);
    \draw [line width=1pt, -{Stealth[scale=1.5]}, ] (20.5,6.75) -- (20.5,7.75);
    \draw [line width=1pt, -{Stealth[scale=1.5]}, ] (20.5,6.75) -- (21.625,6.75);
    \draw [line width=1pt, short] (20.625,6.75) -- (20.75,7.375);
    \draw [line width=1pt, short] (20.75,7.375) -- (21.25,7.375);
    \draw [line width=1pt, short] (21.25,7.375) -- (21.375,6.75);
    \node [font=\fontsize{13.1pt}{17.0pt}\selectfont, fill={rgb,255:red,255; green,255; blue,255}, fill opacity=0, text opacity=1, inner xsep=0.080cm, inner ysep=0.085cm, rounded corners=0.020cm] at (21.25,7.625) {BPF};
    \draw [ line width=1pt ] (20.25,8) rectangle (21.75,6.5);
    \draw [line width=1pt, -{Stealth[scale=1.5]}, ] (21,3.125) -- (21,4.25);
    \draw [line width=1pt, -{Stealth[scale=1.5]}, ] (19,7.125) -- (20.25,7.125);
    \draw [line width=1pt, -{Stealth[scale=1.5]}, ] (21.75,7.25) -- (26.5,7.25);
    \draw [ line width=1pt](26.125,5.125) to[short] (26.75,5.125);
    \draw [line width=1pt, -{Stealth[scale=1.5]}, ] (26.75,5.125) -- (26.75,6.875);
    \draw [line width=1pt, short] (27,7.25) -- (28.125,7.25);
    \draw [ fill={rgb,255:red,255; green,255; blue,255}, fill opacity=0, line width=1pt ] (23.5,6) rectangle (26.25,4.375);
    \node [font=\fontsize{13.1pt}{17.0pt}\selectfont, fill={rgb,255:red,255; green,255; blue,255}, fill opacity=0, text opacity=1, inner xsep=0.080cm, inner ysep=0.085cm, rounded corners=0.020cm, align=center] at (24.875,5.125) {Adaptive FIR \\ Filter};
    \draw [line width=1pt, short] (23.75,4.125) -- (28.125,4.125);
    \draw [line width=1pt, short] (28.125,7.25) -- (28.125,4.125);
    \draw [line width=1pt, -{Stealth[scale=1.5]}, ] (21.75,5.125) -- (23.5,5.125);
    \draw [line width=1pt, short] (27.125,1.25) -- (27.125,-0.125);
    \draw [line width=1pt, short] (27.125,-0.125) -- (2.25,-0.125);
    \node [font=\fontsize{15.9pt}{20.7pt}\selectfont, fill={rgb,255:red,255; green,255; blue,255}, fill opacity=0, text opacity=1, inner xsep=0.080cm, inner ysep=0.085cm, rounded corners=0.020cm] at (21.875,1.125) {\textbf{MCU}};
    \node [font=\fontsize{13.1pt}{17.0pt}\selectfont, fill={rgb,255:red,255; green,255; blue,255}, fill opacity=0, text opacity=1, inner xsep=0.080cm, inner ysep=0.085cm, rounded corners=0.020cm] at (18.25,2.875) {$I[n]$};
    \node [font=\fontsize{13.1pt}{17.0pt}\selectfont, fill={rgb,255:red,255; green,255; blue,255}, fill opacity=0, text opacity=1, inner xsep=0.080cm, inner ysep=0.085cm, rounded corners=0.020cm] at (18.125,2) {$Q[n]$};
    \node [font=\fontsize{13.1pt}{17.0pt}\selectfont, fill={rgb,255:red,255; green,255; blue,255}, fill opacity=0, text opacity=1, inner xsep=0.080cm, inner ysep=0.085cm, rounded corners=0.020cm] at (18.125,7.125) {$x[n]$};
    \node [font=\fontsize{13.1pt}{17.0pt}\selectfont, fill={rgb,255:red,255; green,255; blue,255}, fill opacity=0, text opacity=1, inner xsep=0.080cm, inner ysep=0.085cm, rounded corners=0.020cm] at (19.5,9.875) {\textbf{BLE Module}};
    \draw [line width=1pt, short] (26.125,1.25) -- (26.125,0.25);
    \draw [line width=1pt, short] (26.125,0.25) -- (6.875,0.25);
    \draw [ line width=1pt](6,1) to[short] (6,0.25);
    \draw [ line width=1pt](6.875,0.25) to[short] (6,0.25);
    \draw [ line width=1pt](6,2.875) to[short] (6,3.625);
    \draw [ line width=1pt](6,3.625) to[short] (5.5,3.625);
    \draw [ line width=1pt](9,2.75) to[short] (5.375,2.75);
    \draw [ line width=1pt](5,1.875) to[short] (5,-0.125);
    \draw [ color={rgb,255:red,212; green,0; blue,0}, draw opacity=0.35, line width=1pt, dashed] (3.25,6.125) -- (3.25,0.875);
    \draw [ color={rgb,255:red,212; green,0; blue,0}, draw opacity=0.35, line width=1pt, dashed] (3.25,0.875) -- (6.625,0.875);
    \draw [ color={rgb,255:red,212; green,0; blue,0}, draw opacity=0.35, line width=1pt, dashed] (6.625,1.25) -- (6.625,0.875);
    \draw [ color={rgb,255:red,212; green,0; blue,0}, draw opacity=0.35, line width=1pt, dashed] (3.25,6.125) -- (3.75,6.125);
    \node [font=\fontsize{13.1pt}{17.0pt}\selectfont, fill={rgb,255:red,255; green,255; blue,255}, fill opacity=0, text opacity=1, inner xsep=0.080cm, inner ysep=0.085cm, rounded corners=0.020cm] at (0.375,9.875) {$i+1^{th}$};
    \draw [ color={rgb,255:red,0; green,0; blue,0}, draw opacity=0.27, line width=1pt, dashed] (4.5,0.875) -- (4.5,-0.125);
    \draw [ color={rgb,255:red,0; green,0; blue,0}, draw opacity=0.27, line width=1pt, dashed] (5.75,0.875) -- (5.75,0.25);
    \draw [ color={rgb,255:red,0; green,0; blue,0}, draw opacity=0.59, line width=1pt, dashed] (5.75,0.25) -- (6.125,0.25);
    \draw [ color={rgb,255:red,0; green,0; blue,0}, draw opacity=0.31, line width=1pt, dashed] (6.875,4.875) -- (9,4.875);
    \draw [ color={rgb,255:red,0; green,0; blue,0}, draw opacity=0.31, line width=1pt, dashed] (6.875,4) -- (8.75,4);
    \draw [line width=1pt, -{Stealth[scale=1.5]}, ] (28.125,10) -- (23.25,10);
    \draw [line width=1pt, short] (28.125,7.25) -- (28.125,10);
    \draw [line width=0.8pt](-0.875,1.25) to[C] (-2.125,1.25);
    \draw [ line width=0.8pt](-2.25,-0.25) to[R] (-2.25,1.25);
    \draw [ line width=1pt](-3.125,1.25) to[short] (-2.125,1.25);
    \draw [line width=1pt](-2.25,0) to (-2.25,-0.125) node[sground]{};
    \draw [ line width=1pt](-0.875,1.25) to[short] (0.25,1.25);
    \draw [line width=1pt, short] (14.875,2.625) -- (16.75,2.625);
    \draw [line width=1pt, short] (16.75,2) -- (14.875,2);
    \draw [line width=1pt, short] (14.875,1.375) -- (16.75,1.375);
    \draw [line width=1pt, -{Stealth[scale=1.5]}, ] (16.25,2.625) -- (16.75,2.625);
    \draw [line width=1pt, -{Stealth[scale=1.5]}, ] (16.125,2) -- (16.75,2);
    \draw [line width=1pt, -{Stealth[scale=1.5]}, ] (16,1.375) -- (16.75,1.375);
    \node [font=\fontsize{11.9pt}{15.5pt}\selectfont, fill={rgb,255:red,255; green,255; blue,255}, fill opacity=0, text opacity=1, inner xsep=0.080cm, inner ysep=0.085cm, rounded corners=0.020cm] at (15.375,1) {SPI};
    \node [font=\fontsize{13.1pt}{17.0pt}\selectfont, fill={rgb,255:red,255; green,255; blue,255}, fill opacity=0, text opacity=1, inner xsep=0.080cm, inner ysep=0.085cm, rounded corners=0.020cm, align=center] at (-15.125,-0.625) {ELE embedded \\ Ear Piece Design};
    \node [font=\fontsize{13.1pt}{17.0pt}\selectfont, fill={rgb,255:red,255; green,255; blue,255}, fill opacity=0, text opacity=1, inner xsep=0.080cm, inner ysep=0.085cm, rounded corners=0.020cm] at (-15,6.25) {Electrode Placement};
    \node [font=\fontsize{11.9pt}{15.5pt}\selectfont, fill={rgb,255:red,255; green,255; blue,255}, fill opacity=0, text opacity=1, inner xsep=0.080cm, inner ysep=0.085cm, rounded corners=0.020cm] at (13,10.75) {24-bit $\Sigma$-$\Delta$ ADC};
    \draw [ color={rgb,255:red,0; green,0; blue,0}, draw opacity=0.31, line width=1pt, dashed] (12.125,9.5) -- (12.75,10.125);
    \draw [ color={rgb,255:red,0; green,0; blue,0}, draw opacity=0.31, line width=1pt, dashed] (12.125,9.5) -- (12.75,8.875);
    \draw [ color={rgb,255:red,0; green,0; blue,0}, draw opacity=0.31, line width=1pt, dashed] (12.75,10.125) -- (13.625,10.125);
    \draw [ color={rgb,255:red,0; green,0; blue,0}, draw opacity=0.31, line width=1pt, dashed] (13.625,10.125) -- (13.625,8.875);
    \draw [ color={rgb,255:red,0; green,0; blue,0}, draw opacity=0.31, line width=1pt, dashed] (12.75,8.875) -- (13.625,8.875);
    \draw [line width=1pt, short] (12.125,7) -- (12.75,7.625);
    \draw [line width=1pt, short] (12.75,7.625) -- (13.625,7.625);
    \draw [line width=1pt, short] (12.125,7) -- (12.75,6.375);
    \draw [line width=1pt, short] (12.75,6.375) -- (13.625,6.375);
    \draw [line width=1pt, short] (13.625,7.625) -- (13.625,6.375);
    \draw [line width=1pt, short] (12.75,2.5) -- (13.625,2.5);
    \draw [line width=1pt, short] (13.625,2.5) -- (13.625,1.25);
    \draw [line width=1pt, short] (12.75,1.25) -- (13.625,1.25);
    \draw [line width=1pt, short] (12.125,1.875) -- (12.75,1.25);
    \draw [line width=1pt, short] (12.125,1.875) -- (12.75,2.5);
    \draw [ color={rgb,255:red,0; green,0; blue,0}, draw opacity=0.31, line width=1pt, dashed] (12.75,5.375) -- (13.625,5.375);
    \draw [ color={rgb,255:red,0; green,0; blue,0}, draw opacity=0.31, line width=1pt, dashed] (13.625,5.375) -- (13.625,4.125);
    \draw [ color={rgb,255:red,0; green,0; blue,0}, draw opacity=0.31, line width=1pt, dashed] (12.125,4.75) -- (12.75,5.375);
    \draw [ color={rgb,255:red,0; green,0; blue,0}, draw opacity=0.31, line width=1pt, dashed] (12.125,4) -- (12.75,4.625);
    \draw [ color={rgb,255:red,0; green,0; blue,0}, draw opacity=0.31, line width=1pt, dashed] (12.125,4) -- (12.75,3.375);
    \draw [ color={rgb,255:red,0; green,0; blue,0}, draw opacity=0.31, line width=1pt, dashed] (12.75,3.375) -- (13.625,3.375);
    \draw [ color={rgb,255:red,0; green,0; blue,0}, draw opacity=0.31, line width=1pt, dashed] (13.625,4.625) -- (13.625,3.375);
    \draw [ color={rgb,255:red,0; green,0; blue,0}, draw opacity=0.31, line width=1pt, dashed] (12.75,4.625) -- (13.625,4.625);
    \draw [ color={rgb,255:red,0; green,0; blue,0}, draw opacity=0.31, line width=1pt, dashed] (12.125,4.75) -- (12.5,4.375);
    \draw [line width=1pt, short] (12.125,2.625) -- (12.5,2.25);
    \draw [line width=1pt, short] (12.125,2.625) -- (12.75,3.25);
    \draw [line width=1pt, short] (12.75,3.25) -- (13.625,3.25);
    \draw [line width=1pt, short] (13.625,3.25) -- (13.625,2);
    \draw [ color={rgb,255:red,0; green,0; blue,0}, draw opacity=0.21, line width=1pt, -{Stealth[scale=1.5]}, ] (9,9) -- (10.125,9);
    \draw [ color={rgb,255:red,0; green,0; blue,0}, draw opacity=0.21, line width=1pt, short] (9.5,9.625) -- (9.625,9);
    \draw [ color={rgb,255:red,0; green,0; blue,0}, draw opacity=0.21, line width=1pt, short] (9,9.625) -- (9.5,9.625);
    \draw [ color={rgb,255:red,0; green,0; blue,0}, draw opacity=0.21, line width=1pt, -{Stealth[scale=1.5]}, ] (9,9) -- (9,10);
    \draw [ color={rgb,255:red,0; green,0; blue,0}, draw opacity=0.21 , line width=1pt ] (8.75,10.25) rectangle (10.25,8.75);
    \node [font=\fontsize{13.1pt}{17.0pt}\selectfont, fill={rgb,255:red,255; green,255; blue,255}, fill opacity=0, text opacity=1, inner xsep=0.080cm, inner ysep=0.085cm, rounded corners=0.020cm] at (9.75,7.375) {LPF};
    \draw [line width=1pt, -{Stealth[scale=1.5]}, ] (9,6.5) -- (10.125,6.5);
    \draw [line width=1pt, short] (9.5,7.125) -- (9.625,6.5);
    \draw [line width=1pt, short] (9,7.125) -- (9.5,7.125);
    \draw [line width=1pt, -{Stealth[scale=1.5]}, ] (9,6.5) -- (9,7.5);
    \draw [ line width=1pt ] (8.75,7.75) rectangle (10.25,6.25);
    \node [font=\fontsize{13.1pt}{17.0pt}\selectfont, fill={rgb,255:red,255; green,255; blue,255}, fill opacity=0, text opacity=1, inner xsep=0.080cm, inner ysep=0.085cm, rounded corners=0.020cm] at (9.75,2.25) {LPF};
    \draw [line width=1pt, -{Stealth[scale=1.5]}, ] (9,1.375) -- (10.125,1.375);
    \draw [line width=1pt, short] (9.5,2) -- (9.625,1.375);
    \draw [line width=1pt, short] (9,2) -- (9.5,2);
    \draw [line width=1pt, -{Stealth[scale=1.5]}, ] (9,1.375) -- (9,2.375);
    \draw [ line width=1pt ] (8.75,2.625) rectangle (10.25,1.125);
    \node [font=\fontsize{13.1pt}{17.0pt}\selectfont, fill={rgb,255:red,255; green,255; blue,255}, fill opacity=0, text opacity=1, inner xsep=0.080cm, inner ysep=0.085cm, rounded corners=0.020cm] at (9.75,4.375) {LPF};
    \draw [ color={rgb,255:red,0; green,0; blue,0}, draw opacity=0.21, line width=1pt, -{Stealth[scale=1.5]}, ] (9,3.5) -- (10.125,3.5);
    \draw [ color={rgb,255:red,0; green,0; blue,0}, draw opacity=0.21, line width=1pt, short] (9.5,4.125) -- (9.625,3.5);
    \draw [ color={rgb,255:red,0; green,0; blue,0}, draw opacity=0.21, line width=1pt, short] (9,4.125) -- (9.5,4.125);
    \draw [ color={rgb,255:red,0; green,0; blue,0}, draw opacity=0.21, line width=1pt, -{Stealth[scale=1.5]}, ] (9,3.5) -- (9,4.5);
    \draw [ color={rgb,255:red,0; green,0; blue,0}, draw opacity=0.21 , line width=1pt ] (8.75,4.75) rectangle (10.25,3.25);
    \draw [ color={rgb,255:red,0; green,0; blue,0}, draw opacity=0.21, line width=1pt, short] (9,4.75) -- (9,5.125);
    \draw [ color={rgb,255:red,0; green,0; blue,0}, draw opacity=0.21, line width=1pt, short] (9,5.125) -- (10.5,5.125);
    \draw [ color={rgb,255:red,0; green,0; blue,0}, draw opacity=0.21, line width=1pt, short] (10.5,5.125) -- (10.5,3.625);
    \draw [ color={rgb,255:red,0; green,0; blue,0}, draw opacity=0.21, line width=1pt, short] (10.25,3.625) -- (10.5,3.625);
    \draw [line width=1pt, short] (9,2.625) -- (9,3);
    \draw [line width=1pt, short] (9,3) -- (10.5,3);
    \draw [line width=1pt, short] (10.5,3) -- (10.5,1.5);
    \draw [line width=1pt, short] (10.25,1.5) -- (10.5,1.5);
    \draw [ color={rgb,255:red,0; green,0; blue,0}, draw opacity=0.3, line width=1pt, dashed] (10.25,9.5) -- (12.125,9.5);
    \draw [ line width=1pt](10.25,7) to[short] (12.125,7);
    \draw [ color={rgb,255:red,0; green,0; blue,0}, draw opacity=0.31, line width=1pt, dashed] (10.25,4) -- (12.125,4);
    \draw [ color={rgb,255:red,0; green,0; blue,0}, draw opacity=0.31, line width=1pt, dashed] (10.5,4.75) -- (12.125,4.75);
    \draw [ line width=1pt](10.25,1.875) to[short] (12.125,1.875);
    \draw [ line width=1pt](10.5,2.625) to[short] (12.125,2.625);
    \node [font=\fontsize{11.9pt}{15.5pt}\selectfont, fill={rgb,255:red,255; green,255; blue,255}, fill opacity=0, text opacity=1, inner xsep=0.080cm, inner ysep=0.085cm, rounded corners=0.020cm] at (2.5,9.875) {Raw Bio-Signal};
    \node [font=\fontsize{11.9pt}{15.5pt}\selectfont, fill={rgb,255:red,255; green,255; blue,255}, fill opacity=0, text opacity=1, inner xsep=0.080cm, inner ysep=0.085cm, rounded corners=0.020cm] at (2,7.5) {Raw Bio-Signal};
    \draw [line width=1pt, short] (0.25,1.25) -- (1.75,2.25);
    \draw [line width=1pt, short] (0.25,1.25) -- (1.75,0.25);
    \draw [line width=1pt, short] (1.75,2.25) -- (1.75,0.25);
    \node [font=\fontsize{13.1pt}{17.0pt}\selectfont, fill={rgb,255:red,255; green,255; blue,255}, fill opacity=0, text opacity=1, inner xsep=0.080cm, inner ysep=0.085cm, rounded corners=0.020cm] at (-0.625,0.75) {$1 kHz$};
    \draw [ line width=1pt](1.75,1.25) to[short] (2.25,1.25);
    \draw [line width=1pt, short] (2.25,1.25) -- (2.25,-0.125);
    \draw [-{Stealth[scale=1.5]}, ] (5,3.625) -- (5,3.125);
    \draw [-{Stealth[scale=1.5]}, ] (6,2.75) -- (6,2.25);
    \draw [-{Stealth[scale=1.5]}, ] (5,1.75) -- (5,2.25);
    \draw [-{Stealth[scale=1.5]}, ] (6,1) -- (6,1.5);
    \draw [-{Stealth[scale=1.5]}, ] (8.25,1.875) -- (8.75,1.875);
    \draw [-{Stealth[scale=1.5]}, ] (8.375,2.75) -- (8.875,2.75);
    \draw [ color={rgb,255:red,0; green,0; blue,0}, draw opacity=0.32, -{Stealth[scale=1.5]}, ] (8.375,4) -- (8.75,4);
    \draw [ color={rgb,255:red,0; green,0; blue,0}, draw opacity=0.32, -{Stealth[scale=1.5]}, ] (8.375,4.875) -- (8.8,4.875);
    \draw [ color={rgb,255:red,0; green,0; blue,0}, draw opacity=0.32, -{Stealth[scale=1.5]}, ] (10.625,4.75) -- (11,4.75);
    \draw [ color={rgb,255:red,0; green,0; blue,0}, draw opacity=0.32, -{Stealth[scale=1.5]}, ] (10.625,4) -- (11,4);
    \draw [-{Stealth[scale=1.5]}, ] (10.5,2.625) -- (11,2.625);
    \draw [-{Stealth[scale=1.5]}, ] (10.5,1.875) -- (11,1.875);
    \draw [-{Stealth[scale=1.5]}, ] (10.5,7) -- (11,7);
    \draw [ color={rgb,255:red,0; green,0; blue,0}, draw opacity=0.32, -{Stealth[scale=1.5]}, ] (10.625,9.5) -- (11,9.5);
    \draw [ fill={rgb,255:red,255; green,255; blue,255}, fill opacity=0] (-8.5,12) rectangle (-7.75,11.25);
    \node [font=\fontsize{13.1pt}{17.0pt}\selectfont, fill={rgb,255:red,255; green,255; blue,255}, fill opacity=0, text opacity=1, inner xsep=0.080cm, inner ysep=0.085cm, rounded corners=0.020cm] at (-8.125,11.625) {\textbf{C}};
    \draw [ fill={rgb,255:red,255; green,255; blue,255}, fill opacity=0] (-2.375,11.875) rectangle (-1.625,11.125);
    \node [font=\fontsize{13.1pt}{17.0pt}\selectfont, fill={rgb,255:red,255; green,255; blue,255}, fill opacity=0, text opacity=1, inner xsep=0.080cm, inner ysep=0.085cm, rounded corners=0.020cm] at (-2,11.5) {\textbf{D}};
    \node [font=\fontsize{15.9pt}{20.7pt}\selectfont, fill={rgb,255:red,255; green,255; blue,255}, fill opacity=0, text opacity=1, inner xsep=0.080cm, inner ysep=0.085cm, rounded corners=0.020cm] at (-10.625,3.125) {$V_{CM}$};
    \draw [ fill={rgb,255:red,255; green,255; blue,255}, fill opacity=0] (-0.25,0.375) rectangle (0.5,-0.375);
    \node [font=\fontsize{13.1pt}{17.0pt}\selectfont, fill={rgb,255:red,255; green,255; blue,255}, fill opacity=0, text opacity=1, inner xsep=0.080cm, inner ysep=0.085cm, rounded corners=0.008cm] at (0.125,0) {\textbf{F}};
    \draw [ fill={rgb,255:red,255; green,255; blue,255}, fill opacity=0] (-3,3.5) rectangle (-2.25,2.75);
    \node [font=\fontsize{13.1pt}{17.0pt}\selectfont, fill={rgb,255:red,255; green,255; blue,255}, fill opacity=0, text opacity=1, inner xsep=0.080cm, inner ysep=0.085cm, rounded corners=0.020cm] at (-2.625,3.125) {\textbf{E}};
    \draw [ fill={rgb,255:red,255; green,255; blue,255}, fill opacity=0] (6,6.375) rectangle (6.75,5.625);
    \node [font=\fontsize{13.1pt}{17.0pt}\selectfont, fill={rgb,255:red,255; green,255; blue,255}, fill opacity=0, text opacity=1, inner xsep=0.080cm, inner ysep=0.085cm, rounded corners=0.020cm] at (6.375,6) {\textbf{G}};
    \draw [ fill={rgb,255:red,255; green,255; blue,255}, fill opacity=0] (9.125,11.625) rectangle (9.875,10.875);
    \node [font=\fontsize{13.1pt}{17.0pt}\selectfont, fill={rgb,255:red,255; green,255; blue,255}, fill opacity=0, text opacity=1, inner xsep=0.080cm, inner ysep=0.085cm, rounded corners=0.020cm] at (9.5,11.25) {\textbf{H}};
    \draw [ fill={rgb,255:red,255; green,255; blue,255}, fill opacity=0] (14,1.625) rectangle (14.75,0.875);
    \node [font=\fontsize{13.1pt}{17.0pt}\selectfont, fill={rgb,255:red,255; green,255; blue,255}, fill opacity=0, text opacity=1, inner xsep=0.080cm, inner ysep=0.085cm, rounded corners=0.000cm] at (14.375,1.25) {\textbf{I}};
    \draw [ fill={rgb,255:red,255; green,255; blue,255}, fill opacity=0] (15.875,8.625) rectangle (16.625,7.875);
    \node [font=\fontsize{13.1pt}{17.0pt}\selectfont, fill={rgb,255:red,255; green,255; blue,255}, fill opacity=0, text opacity=1, inner xsep=0.080cm, inner ysep=0.085cm, rounded corners=0.000cm] at (16.25,8.25) {\textbf{J}};
    \draw [ fill={rgb,255:red,255; green,255; blue,255}, fill opacity=0] (24.25,3.75) rectangle (25,3);
    \node [font=\fontsize{13.1pt}{17.0pt}\selectfont, fill={rgb,255:red,255; green,255; blue,255}, fill opacity=0, text opacity=1, inner xsep=0.080cm, inner ysep=0.085cm, rounded corners=0.020cm] at (24.625,3.375) {\textbf{M}};
    \draw [ fill={rgb,255:red,255; green,255; blue,255}, fill opacity=0] (19.25,4.125) rectangle (20,3.375);
    \node [font=\fontsize{13.1pt}{17.0pt}\selectfont, fill={rgb,255:red,255; green,255; blue,255}, fill opacity=0, text opacity=1, inner xsep=0.080cm, inner ysep=0.085cm, rounded corners=0.020cm] at (19.625,3.75) {\textbf{L}};
    \draw [ fill={rgb,255:red,255; green,255; blue,255}, fill opacity=0] (17.75,11.875) rectangle (18.5,11.125);
    \node [font=\fontsize{13.1pt}{17.0pt}\selectfont, fill={rgb,255:red,255; green,255; blue,255}, fill opacity=0, text opacity=1, inner xsep=0.080cm, inner ysep=0.085cm, rounded corners=0.020cm] at (18.125,11.5) {\textbf{K}};
    \draw [line width=1pt, -{Stealth[scale=1.5]}, ] (4.5,5.125) -- (4.5,6.125);
    \node [font=\fontsize{14.8pt}{19.2pt}\selectfont, fill={rgb,255:red,255; green,255; blue,255}, fill opacity=0, text opacity=1, inner xsep=0.080cm, inner ysep=0.085cm, rounded corners=0.020cm] at (-2.375,1.75) {$V_{add}$};
    \node [font=\fontsize{14.8pt}{19.2pt}\selectfont, fill={rgb,255:red,255; green,255; blue,255}, fill opacity=0, text opacity=1, inner xsep=0.080cm, inner ysep=0.085cm, rounded corners=0.020cm] at (5.625,4.125) {$V_{BP,i}$};
    \draw [ fill={rgb,255:red,255; green,255; blue,255}, fill opacity=0] (-16.875,11.75) rectangle (-16.125,11);
    \node [font=\fontsize{13.1pt}{17.0pt}\selectfont, fill={rgb,255:red,255; green,255; blue,255}, fill opacity=0, text opacity=1, inner xsep=0.080cm, inner ysep=0.085cm, rounded corners=0.020cm] at (-16.5,11.375) {\textbf{A}};
    \draw [-{Stealth[scale=1.5]}, ] (8.125,7) -- (8.625,7);

    \draw [ color={rgb,255:red,0; green,0; blue,0}, draw opacity=0.32, -{Stealth[scale=1.5]}, ] (8.25,9.5) -- (8.625,9.5);

    \node[inner sep=0] at (-15.125, 2.8) {\includegraphics[width=3cm]{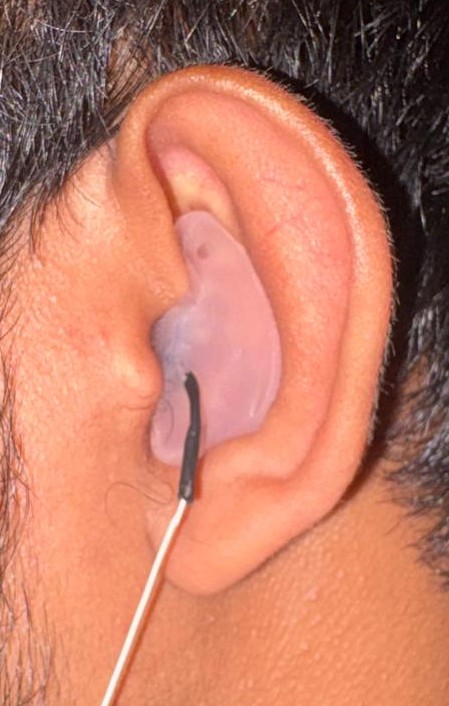}};
    \node[inner sep=0] at (-15.125, 8.8) {\includegraphics[width=3cm]{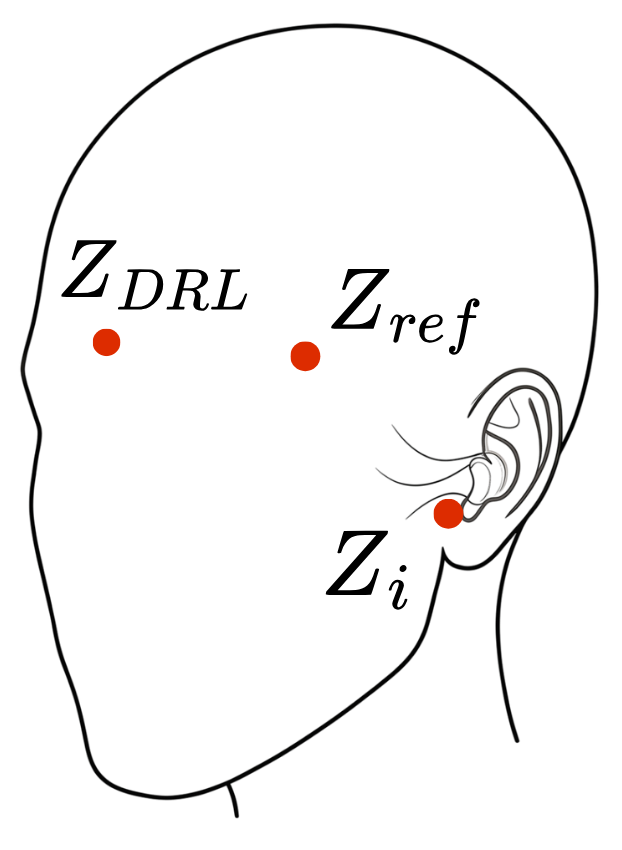}};
    \node[inner sep=0] at (21.875,9.875) {\includegraphics[width=1cm]{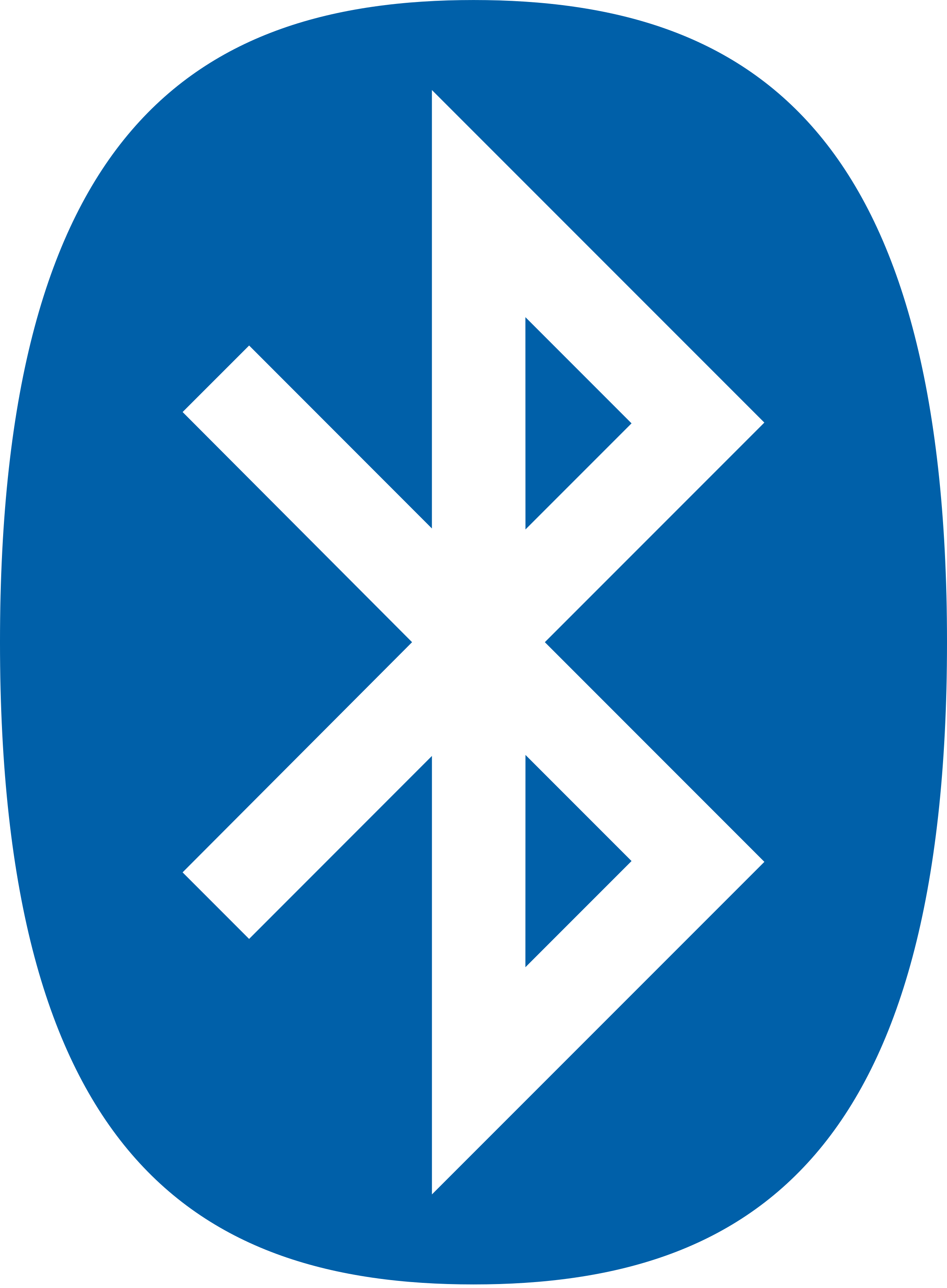}};

    \draw [ fill={rgb,255:red,255; green,255; blue,255}, fill opacity=1] (-16.875,5.625) rectangle (-16.125,4.875);
    \node [font=\fontsize{13.1pt}{17.0pt}\selectfont, fill={rgb,255:red,255; green,255; blue,255}, fill opacity=1, text opacity=1, inner xsep=0.080cm, inner ysep=0.085cm, rounded corners=0.020cm] at (-16.5,5.25) {\textbf{B}};
    \end{circuitikz}
    
}%

%% file: ref.bib
@article{kappel2017physiological,
  title={Physiological artifacts in scalp {EEG} and {Ear}-{EEG}},
  author={Kappel, Simon L. and Looney, David and Mandic, Danilo P. and Kidmose, Preben},
  journal={Biomedical Engineering Online},
  volume={16},
  number={1},
  pages={103},
  year={2017},
  publisher={Springer},
  doi={10.1186/s12938-017-0391-2}
}

@article{mikkelsen_eeg_2015,
  title={{EEG} Recorded from the Ear: Characterizing the {Ear}-{EEG} Method},
  author={Mikkelsen, Kaare B. and Kappel, Simon L. and Mandic, Danilo P. and Kidmose, Preben},
  journal={Frontiers in Neuroscience},
  volume={9},
  pages={438},
  year={2015},
  doi={10.3389/fnins.2015.00438},
}

@article{giangrande_motion_2024,
  title={Motion Artifacts in Dynamic {EEG} Recordings: Experimental Observations, Electrical Modelling, and Design Considerations},
  author={Giangrande, Alessandra and Botter, Alberto and Piitulainen, Harri and Cerone, Giacinto Luigi},
  journal={Sensors},
  volume={24},
  number={19},
  pages={6363},
  year={2024},
  doi={10.3390/s24196363},

}

@inproceedings{10708999,
  author={Nan, Jiedong},
  booktitle={IEEE 2nd International Conference on Image Processing and Computer Applications},
  title={Comparative Study of Motion Artifact Removal Algorithms},
  year={2024},
  pages={1178--1187},
  doi={10.1109/ICIPCA61593.2024.10708999}
}

@article{s22082948,
  author={Phadikar, Souvik and Sinha, Nidul and Ghosh, Rajdeep and Ghaderpour, Ebrahim},
  title={Automatic Muscle Artifacts Identification and Removal from Single-Channel {EEG} Using Wavelet Transform with Meta-Heuristically Optimized Non-Local Means Filter},
  journal={Sensors},
  volume={22},
  number={8},
  article-number={2948},
  year={2022},
  doi={10.3390/s22082948},
}

@article{stalin2021machine,
  title={A Machine Learning-Based Big {EEG} Data Artifact Detection and Wavelet-Based Removal: An Empirical Approach},
  author={Stalin, Shalini and Roy, Vandana and Shukla, Prashant Kumar and Zaguia, Atef and Khan, Mohammad Monirujjaman and Shukla, Piyush Kumar and Jain, Anurag},
  journal={Mathematical Problems in Engineering},
  volume={2021},
  number={1},
  year={2021},
  publisher={Wiley Online Library},
  doi={10.1155/2021/2942808}
}

@inproceedings{inproceedings,
  author={Hossain, Md Shafayet and Mahmud, Sakib and Khandakar, Amith},
  title={Deep Learning Technique to Denoise Electromyogram Artifacts from Single-Channel Electroencephalogram Signals},
  booktitle={Proceedings of the International Conference on Computer Science and Information Technology},
  pages={85--98},
  year={2022},
  doi={10.5121/csit.2022.122006}
}

@article{10.3389/fnins.2021.611962,
  author={Rosanne, Olivier and Albuquerque, Isabela and Cassani, Raymundo and Gagnon, Jean-François and Tremblay, Sebastien and Falk, Tiago H.},
  title={Adaptive Filtering for Improved {EEG}-Based Mental Workload Assessment of Ambulant Users},
  journal={Frontiers in Neuroscience},
  volume={15},
  year={2021},
  doi={10.3389/fnins.2021.611962},
}

@article{10286273,
  author={Yadav, Shubham and Saha, Suman Kumar and Kar, Rajib},
  title={Evolutionary Algorithm-Based Optimal Wiener-Adaptive Filter Design: An Application on {EEG} Noise Mitigation},
  journal={IEEE Transactions on Instrumentation and Measurement},
  volume={72},
  pages={1--12},
  year={2023},
  doi={10.1109/TIM.2023.3324345}
}

@inproceedings{degen2007improved,
  title={An Improved Method to Continuously Monitor the Electrode-Skin Impedance During Bioelectric Measurements},
  author={Degen, Thomas and Loeliger, Teddy},
  booktitle={29th Annual International Conference of the IEEE Engineering in Medicine and Biology Society},
  pages={6294--6297},
  year={2007},
  organization={IEEE},
  doi={10.1109/IEMBS.2007.4353760}
}

@article{degen2008continuous,
  title={Continuous Monitoring of Electrode--Skin Impedance Mismatch During Bioelectric Recordings},
  author={Degen, Thomas and J{\"a}ckel, Heinz},
  journal={IEEE Transactions on Biomedical Engineering},
  volume={55},
  number={6},
  pages={1711--1715},
  year={2008},
  publisher={IEEE},
  doi={10.1109/TBME.2008.919120}
}

@inbook{inbook,
  author    = {Diniz, Paulo S. R.},
  year      = {2008},
  month     = {},
  pages     = {},
  title     = {Adaptive Filtering: Algorithms and Practical Implementation},
  publisher = {Springer},
  address   = {New York, NY, USA},
  doi       = {10.1007/978-0-387-68606-6}
}

@article{ghaleb2018two,
  author  = {F. A. Ghaleb and M. B. Kamat and M. Salleh and M. F. Rohani and S. A. Razak},
  title   = {Two-stage motion artefact reduction algorithm for electrocardiogram using weighted adaptive noise cancelling and recursive {Hampel} filter},
  journal = {PLOS ONE},
  volume  = {13},
  number  = {11},
  month   = nov,
  year    = {2018},
  doi     = {10.1371/journal.pone.0207176}
}

@INPROCEEDINGS{muller,
  author={Pandey, Aviral and Alamouti, Sina Faraji and Doong, Justin and Kaveh, Ryan and Yalcin, Cem and Ghanbari, Mohammad Meraj and Muller, Rikky},
  booktitle={IEEE Custom Integrated Circuits Conference}, 
  title={A 6.8µW {AFE} for {Ear EEG} Recording with Simultaneous Impedance Measurement for Motion Artifact Cancellation}, 
  year={2022},
  volume={},
  number={},
  pages={1-2},
  doi={10.1109/CICC53496.2022.9772839}}

@phdthesis{Kaveh:EECS-2023-56,
    Author= {Kaveh, Ryan},
    Title= {{Ear EEG}: Sensors and Systems for User-generic Neural Hearables},
    School= {EECS Department, University of California, Berkeley},
    Year= {2023},
    Month= {May},

    Number= {UCB/EECS-2023-56},

}

@ARTICLE{10.3389/fnins.2024.1441897,
    
AUTHOR={Moumane, Hanane  and Pazuelo, Jérémy  and Nassar, Mérie  and Juez, Jose Yesith  and Valderrama, Mario  and Le Van Quyen, Michel },
           
TITLE={Signal quality evaluation of an in-ear {EEG} device in comparison to a conventional cap system},
          
JOURNAL={Frontiers in Neuroscience},
          
VOLUME={18},
  
YEAR={2024},
 
DOI={10.3389/fnins.2024.1441897},
  
ISSN={1662-453X},
}

@ARTICLE{1161901,
  author={Welch, P.},
  journal={IEEE Transactions on Audio and Electroacoustics}, 
  title={The use of fast Fourier transform for the estimation of power spectra: A method based on time averaging over short, modified periodograms}, 
  year={1967},
  volume={15},
  number={2},
  pages={70-73},
  doi={10.1109/TAU.1967.1161901}}

@article{mikkelsen2019earEEGsleep,
  author  = {Mikkelsen, Kaare B. and Tabar, Yousef Rezaei and Kappel, Simon L. and Christensen, Christoffer B. and Toft, Henrik O. and Hemmsen, Martin C. and Rank, Michelle L. and Otto, Morten and Kidmose, Preben},
  title   = {Accurate whole-night sleep monitoring with dry-contact {Ear-EEG}},
  journal = {Scientific Reports},
  volume  = {9},
  number  = {1},
  pages   = {16824},
  year    = {2019},
  month   = nov,
  doi     = {10.1038/s41598-019-53115-3}
}

@article{ZIBRANDTSEN20172454,
title = {{Ear-EEG} detects ictal and interictal abnormalities in focal and generalized epilepsy – A comparison with scalp {EEG} monitoring},
journal = {Clinical Neurophysiology},
volume = {128},
number = {12},
pages = {2454-2461},
year = {2017},
issn = {1388-2457},
doi = {https://doi.org/10.1016/j.clinph.2017.09.115},
author = {I.C. Zibrandtsen and P. Kidmose and C.B. Christensen and T.W. Kjaer},
}

@patent{azemi2023biosignal,
  author       = {Azemi, Erdrin and Moin, Ali and Pragada, Anuranjini and Lu, Jean Hsiang-Chun and Powell, Victoria M. and Minxha, Juri and Hotelling, Steven P.},
  title        = {Biosignal Sensing Device Using Dynamic Selection of Electrodes},
  number       = {US20230225659A1},
  type         = {Patent Application Publication},
  holder       = {Apple Inc.},
  location     = {United States},
  year         = {2023},
  month        = jul,
  day          = {20},

}

@ARTICLE{chi2010,
  author={Chi, Yu Mike and Jung, Tzyy-Ping and Cauwenberghs, Gert},
  journal={IEEE Reviews in Biomedical Engineering}, 
  title={Dry-Contact and Noncontact Biopotential Electrodes: Methodological Review}, 
  year={2010},
  volume={3},
  number={},
  pages={106-119},
  doi={10.1109/RBME.2010.2084078}}

@ARTICLE{cheon2024,
  author={Cheon, Song-I and Choi, Haidam and Kang, Hyoju and Suh, Ji-Hoon and Park, Seonghyun and Kweon, Soon-Jae and Je, Minkyu and Ha, Sohmyung},
  journal={IEEE Transactions on Biomedical Circuits and Systems}, 
  title={Impedance-Readout Integrated Circuits for Electrical Impedance Spectroscopy: Methodological Review}, 
  year={2024},
  volume={18},
  number={1},
  pages={215-232},
  doi={10.1109/TBCAS.2023.3319212}}
